\documentclass[12pt]{article}

\usepackage[dvipsname]{xcolor}

\usepackage{graphicx}
\usepackage{amsmath}        
\usepackage{amssymb}        
\usepackage{slashed}
\usepackage{bm}
\usepackage{here}
\usepackage{cite}

\def\mydate{November 30, 2023}
\def\ignore#1{{}}

\def\go{\rightarrow}
\def\dd{\partial}

\def\eff{{\rm eff}}
\def\SM{{\rm SM}}
\def\KK{{\rm KK}}

\def\GHU{{\rm GHU}}

\def\onehalf{\hbox{$\frac{1}{2}$}}

\def\la{\langle}
\def\ra{\rangle}

\def\mybig{\displaystyle \strut }

\def\myfrac#1#2{\frac{\mybig #1}{\mybig #2}}

\def\mymat#1#2{\begin{matrix}#1 \cr \noalign{\kern -2pt} #2\end{matrix}}

\def\mynoalign{\noalign{\kern 4pt}}
\def\mysnoalign{\noalign{\kern 3pt}}
\def\mytinynoalign{\noalign{\kern 2pt}}
\def\ignore#1{{}}

\makeatletter
\@addtoreset{equation}{section}
\makeatother

\begin{document}

\thispagestyle{empty}

{\small \noindent \mydate \hfill}

\vskip 2.0cm

\baselineskip=30pt plus 1pt minus 1pt

\begin{center}
{\bf \LARGE $W$ boson mass  in gauge-Higgs unification}\\ 
\end{center}


\baselineskip=22pt plus 1pt minus 1pt

\vskip 1.5cm

\begin{center}
\renewcommand{\thefootnote}{\fnsymbol{footnote}}
{\bf  Yutaka Hosotani$^{a}$\footnote[1]{hosotani@rcnp.osaka-u.ac.jp},
Shuichiro Funatsu$^b$, Hisaki Hatanaka$^c$}

{\bf  Yuta Orikasa$^d$ and Naoki Yamatsu$^e$}



\baselineskip=18pt plus 1pt minus 1pt

\vskip 10pt
{\small \it $^a$Research Center for Nuclear Physics, Osaka University,  Ibaraki, Osaka 567-0047, Japan}\\
{\small \it $^b$Ushiku, Ibaraki 300-1234, Japan} \\
{\small \it $^c$Osaka, Osaka 536-0014, Japan} \\
{\small \it $^d$Institute of Experimental and Applied Physics, Czech Technical University in Prague,} \\
{\small \it Husova 240/5, 110 00 Prague 1, Czech Republic} \\
{\small \it $^e$Department of Physics, National Taiwan University,Taipei, Taiwan 10617, R.O.C.} \\

\end{center}

\vskip 2.cm
\baselineskip=18pt plus 1pt minus 1pt

\begin{abstract}
The $W$ boson mass $m_W$ in the GUT inspired $SO(5) \times U(1) \times SU(3)$ gauge-Higgs unification 
in the Randall-Sundrum (RS) warped  space is evaluated.
The muon decay $\mu^- \go e^- \bar{\nu}_e \nu_\mu$ proceeds by the exchange of not only  the zero mode of the $W$ boson
 $(W^{(0)}$)
but also Kaluza-Klein (KK) excited modes $W^{(n)}$ and $W_R^{(n)}$ ($n \ge 1$) at the tree level.
The anti-de Sitter curvature of  the RS space also affects the relationship among the gauge couplings and 
the  ratio of $m_W$ to the $Z$ boson mass $m_Z$.  The $W$ couplings of leptons and quarks also change.
With  the given KK mass scale $m_\KK$ the range of the Aharonov-Bohm phase $\theta_H$ 
in the fifth dimension is constrained.  
For $m_\KK = 13\,$TeV,  $0.085 \lesssim \theta_H \lesssim 0.11$ and 
$80.381\,{\rm GeV}  \lesssim m_W \lesssim 80.407\,{\rm GeV}$.
The predicted value of $m_W$ for $13\, {\rm TeV} \le m_\KK \le 20\, {\rm TeV}$
lies between $m_W^{\rm SM} = 80.354 \pm 0.007\,$GeV in the standard model and
$m_W^{\rm CDF} = 80.4335 \pm 0.0094\,$GeV, the value reported by the CDF collaboration in 2022.
\end{abstract}


\newpage

\baselineskip=20pt plus 1pt minus 1pt
\parskip=0pt

\section{Introduction} 

Last year the CDF collaboration reported   on the mass of the $W$ boson, 
$m_W^{\rm CDF} = 80.4335 \pm 0.0094\,$GeV.\cite{CDF2022}
The predicted value in the standard model (SM) is $m_W^{\rm SM} =  80.354 \pm 0.007\,$GeV.\cite{Haller2022, Awramik2021,deBlas2022}
The discrepancy between the two has triggered huge debates on possible new physics beyond the SM.
The ATLAS collaboration also reanalyzed the data in 2011 to obtain 
$m_W^{\rm ATLAS} = 80.360 \pm 0.016\,$GeV.\cite{ATLAS2023}
Although the experimental situation  has not been settled yet, it is worth to examine 
various models to find whether or not they can lead to a larger value for $m_W$ than $m_W^{\rm SM} $
without conflicting with other observation at low energies.

There have been various proposals to account for the $m_W$ anomaly.  
Many of them are based on new physics effects on the Peskin-Takeuchi oblique $S$ and $T$ parameters
\cite{Lu2022, Strumia2022, Cacciapaglia2022, SMEFT2022, Asadi2023}
either at the tree level \cite{ Babu2022, Kanemura2022} or 
at the loop level.\cite{Lee2022, Kawamura2022, Nagao2023} 
Another approach is based on a scenario in which new fields and/or couplings give additional
contributions to the Fermi constant $G_F$.\cite{Mishima2022, Blennow2022}

It has been known that  the SM, $SU(3)_C \times SU(2)_L \times U(1)_Y$ gauge theory,
though being mostly successful in describing  phenomena at low energies,  has a severe gauge hierarchy 
problem when embedded in a larger theory such as grand unification.
As one possible answer to this problem  the gauge-Higgs unification  (GHU) scenario has been proposed in which 
gauge symmetry is dynamically broken by an Aharonov-Bohm (AB)  phase, $\theta_H$, in the fifth dimension.
The 4D Higgs boson appears as a 4D fluctuation mode of 
$\theta_H$.
\cite{Hosotani1983, Davies1988, Hosotani1989, Davies1989,  Hatanaka1998, Hatanaka1999,  
Kubo2002, Scrucca2003, ACP2005, Cacciapaglia2006, Medina2007, HOOS2008, FHHOS2013, Yoon2018b,
GUTinspired2019a, FCNC2020a, GUTinspired2020b}

Among various GHU models the 
$SO(5)\times U(1) \times SU(3)$ GHU in the Randall-Sundrum (RS) 
warped  space, inspired from the $SO(11)$ gauge-Higgs grand unification model \cite{SO11GHGU},
has been extensively investigated.\cite{GUTinspired2019a, FCNC2020a, GUTinspired2020b}   
It has been shown that the grand unified theory (GUT) inspired GHU yields
nearly the same phenomenology at low energies as the SM.
GHU models in the RS warped space predict, in general, large parity violation in the couplings of quarks and leptons
to Kaluza-Klein (KK) excited modes of gauge bosons, which can be clearly seen, for $m_\KK \sim 13\,$TeV,
for instance, at electron-positron ($e^- e^+$) colliders 
such as ILC with $\sqrt{s} = 250\,$GeV by using polarized $e^-$ and $e^+$ 
beams.\cite{Funatsu2017a, Yoon2018a, Funatsu2019a, GUTinspired2020c, Funatsu2022b}
Deviation from the SM can be explored also in the processes of $W^- W^+$ production and single Higgs production
in $e^- e^+$ collisions.\cite{Funatsu2023a, Yamatsu2023}
Signals of $Z'$ particles, namely KK excited modes of $\gamma$, $Z$, and $Z_R$ gauge bosons,
should be seen in high-luminosity LHC as well.\cite{Funatsu2022a}

KK modes of fermions and gauge bosons in the RS warped space have quite nontrivial couplings.
Recently oblique corrections to $\gamma$, $Z$ and $W$ propagators at the one loop level in the GUT inspired
GHU have been evaluated.\cite{sumrule2023}  Inside the loops all possible KK modes of fermions run.  
The total oblique corrections to $S$, $T$ and $U$ turned out small as a consequence of the coupling sum rules,
special relations holding among infinitely many gauge couplings in the KK mode space.
In the GHU scenario the 5D gauge invariance seems to lead to many surprising coupling relations among 
zero modes and  KK excited modes of fermions and gauge bosons.
It has been shown in the $SU(2)$ GHU model in the RS space that gauge anomalies associated with various 4D modes 
of gauge fields vary with the AB phase $\theta_H$.  The total gauge anomalies obtained by summing contributions from
all fermion KK modes are expressed in terms of the values of the gauge field wave functions at the UV 
and IR branes of the  RS space,  representing relations among gauge couplings of right-handed and left-handed 
fermions.\cite{AnomalyFlow1, AnomalyFlow2}

In view of these facts one may ask how large the $W$ boson mass  $m_W$  is in the GUT inspired GHU.
There are KK excited modes of $W$ and $W_R$ gauge bosons which couple to leptons and contribute
to the muon decay at the tree level.  Further the relation between the gauge couplings and the ratio of
the $W$ and $Z$ boson masses, $m_W/m_Z$,  is changed  even at the tree level.
In this paper we  analyze  this matter in detail.
Additional relevant parameters in the GUT inspired GHU are the KK mass scale $m_\KK$ and the AB phase $\theta_H$.
It will be seen below that for $m_\KK = 13\,$TeV, for instance, $0.085 \lesssim \theta_H \lesssim 0.11$ is
allowed and $m_W$ becomes $80.381 \,{\rm GeV}  \lesssim m_W \lesssim 80.407\,{\rm GeV}$.
The dominant contributions come from large gauge couplings of left-handed leptons to the first KK excited mode 
of the $W$ boson, $W^{(1)}$,  the change in the $W$ couplings of leptons ($e$ and $\mu$), and 
the change in the relation between the gauge couplings and mass ratio $m_W/m_Z$
in the RS warped space.

In Section 2 the  GUT inspired $SO(5) \times U(1)_X \times SU(3)_C$ GHU model is explained.
In Section 3 the $W$ boson mass is evaluated.  It will be seen that the predicted value of $m_W$ in 
the GUT inspired GHU is mostly determined by the value of $\theta_H$.
A summary is given in Section 4.

\section{GUT inspired GHU} 

The GUT inspired $SO(5) \times U(1)_X \times SU(3)_C$ GHU was introduced in ref.\ \cite{GUTinspired2019a}.
It is defined in the RS warped space whose metric is given by\cite{RS1}
\begin{align}
ds^2= g_{MN} dx^M dx^N =
e^{-2\sigma(y)} \eta_{\mu\nu}dx^\mu dx^\nu+dy^2,
\label{RSmetric1}
\end{align}
where $M,N=0,1,2,3,5$, $\mu,\nu=0,1,2,3$, $y=x^5$, $\eta_{\mu\nu}=\mbox{diag}(-1,+1,+1,+1)$,
$\sigma(y)=\sigma(y+ 2L)=\sigma(-y)$, and $\sigma(y)=ky$ for $0 \le y \le L$.
In terms of the conformal coordinate $z=e^{ky}$ ($0 \le y \le L$, $1\leq z\leq z_L=e^{kL}$)
\begin{align}
ds^2=  \frac{1}{z^2} \bigg(\eta_{\mu\nu}dx^{\mu} dx^{\nu} + \frac{dz^2}{k^2}\bigg)~ .
\label{RSmetric-2}
\end{align}
The bulk region $0<y<L$ is anti-de Sitter (AdS) spacetime 
with a cosmological constant $\Lambda=-6k^2$, which is sandwiched by the
UV brane at $y=0$  and the IR brane at $y=L$.  
$z_L$ is called as the warp factor.
The KK mass scale is given by $m_{\rm KK}=\pi k/(z_L-1) \simeq \pi kz_L^{-1}$ for $z_L\gg 1$.

Gauge fields 
$A_M^{SO(5)}$,  $A_M^{U(1)_X}$ and $A_M^{SU(3)_C}$ of $SO(5) \times U(1)_X \times SU(3)_C$
satisfy the orbifold boundary conditions (BCs)
\begin{align}
&\begin{pmatrix} A_\mu \cr  A_{y} \end{pmatrix} (x,y_j-y) =
P_{j} \begin{pmatrix} A_\mu \cr  - A_{y} \end{pmatrix} (x,y_j+y)P_{j}^{-1}
\quad (j=0,1)
\label{BC-gauge1}
\end{align}
where $(y_0, y_1) = (0, L)$.  
Here $P_0=P_1 = P_{\bf 5}^{SO(5)} =\mbox{diag} (I_{4},-I_{1} )$ for $A_M^{SO(5)}$ in the vector representation and 
$P_0=P_1= 1$ for $A_M^{U(1)_X}$ and $A_M^{SU(3)_C}$.
The 4D Higgs field is contained in the $SO(5)/SO(4)$ part of $A_y^{SO(5)}$.
The orbifold BCs break $SO(5)$ to $SO(4) \simeq SU(2)_L \times SU(2)_R$.

The matter content in the GUT inspired GHU is summarized in Table \ref{Table:matter}.
Quark and lepton multiplets are introduced in three generations.  
The lepton multiplets $\Psi_{({\bf 1,4})}^\alpha (x,y)$ ($\alpha=1,2,3$) satisfy BCs
\begin{align}
\Psi_{({\bf 1,4})}^\alpha (x, y_j - y) = 
- P_{\bf 4}^{SO(5)} \gamma^5 \Psi_{({\bf 1,4})}^\alpha (x, y_j + y) 
\label{leptonBC}
\end{align}
where $P_{\bf 4}^{SO(5)} =\mbox{diag} (I_{2},-I_{2} )$.
(For BCs of other multiplets, see ref.\  \cite{GUTinspired2019a} or \cite{sumrule2023}.)
The action of $\Psi_{({\bf 1,4})}^\alpha$ in the bulk is
\begin{align}
&S_{\rm bulk}^{\rm lepton} =  \int d^5x\sqrt{-\det G} \,
 \sum_\alpha  \overline{\Psi}_{({\bf 1,4})}^\alpha {\cal D} (c_\alpha) \Psi_{({\bf 1,4})}^\alpha ~,   \cr
\noalign{\kern 3pt}
&{\cal D}(c)= \gamma^A {e_A}^M
\bigg( D_M+\frac{1}{8}\omega_{MBC}[\gamma^B,\gamma^C]  \bigg) -c\sigma'(y) ~, \cr
\noalign{\kern 3pt}
&D_M =  \dd_M   -i g_A A_M^{SO(5)}  -i g_B Q_X A_M ^{U(1)} ~. 
\label{fermionAction1}
\end{align} 
The dimensionless parameter $c$ in ${\cal D}(c)$ is called the bulk mass parameter, which controls
the wave functions of the zero modes of the fermions.  In the GUT inspired GHU  the bulk mass parameters
are negative for both lepton and quark multiplets.
On the UV brane at $y=0$, gauge-singlet Majorana fermions ${\chi}_{({\bf 1,1})}^\alpha$ and 
one brane scalar  $\Phi_{({\bf 1,4})}$  are introduced.   
There arise gauge-invariant brane interactions of the form 
$\big\{ \tilde{\kappa}_{\bf 1}^{\alpha \beta} \,
\overline{\chi}_{({\bf 1,1})}^\beta 
\tilde{\Phi} {}_{({\bf 1,4})}^\dag \Psi_{({\bf 1,4})}^{\alpha}   + {\rm h.c.} \big\} \delta(y)$
where $\tilde{\Phi}_{({\bf 1,4})}$  denotes a conjugate field in $({\bf 1,4})$
formed from $\Phi_{({\bf 1,4})}^*$.  
The brane scalar field $\Phi_{({\bf 1,4})}$ spontaneously develops a nonvanishing expectation value $\la  \Phi \ra \not= 0$, 
which, with the brane interaction term, induces the inverse seesaw mechanism for neutrinos.

\begin{table}[tbh]
{
\renewcommand{\arraystretch}{1.5}
\begin{center}
\caption{The matter fields in the GUT inspired $SO(5)\times U(1) \times SU(3)$ gauge-Higgs unification.
$(SU(3)_C, SO(5))_{U(1)_X}$ content of each field is shown in the last column.  }
\vskip 10pt
\begin{tabular}{|c|c|c|}
\hline
in the bulk &quark
&$({\bf 3}, {\bf 4})_{\frac{1}{6}} ~ ({\bf 3}, {\bf 1})_{-\frac{1}{3}}^+ 
    ~ ({\bf 3}, {\bf 1})_{-\frac{1}{3}}^-$\\
\cline{2-3}
&lepton
&$\strut ({\bf 1}, {\bf 4})_{-\frac{1}{2}}$ \\
\cline{2-3}
&dark fermion $\Psi^D$ & $({\bf 3}, {\bf 4})_{\frac{1}{6}} ~ ({\bf 1}, {\bf 5})_{0}^+ ~ ({\bf 1}, {\bf 5})_{0}^-$  \\
\hline 
on the UV brane
&Majorana fermion $ \chi$ &$({\bf 1}, {\bf 1})_{0} $ \\
\cline{2-3}
&brane scalar $\Phi$ &$({\bf 1}, {\bf 4})_{\frac{1}{2}} $ \\
\hline
\end{tabular}
\label{Table:matter}
\end{center}
}
\end{table}

In the electroweak sector there are two 5D gauge couplings, $g_A$ and $g_B$, corresponding to
the gauge groups $SO(5)$ and $U(1)_X$, respectively.
The 5D gauge coupling $g_Y^{\rm 5D}$ of $U(1)_{Y}$  is given by $g_Y^{\rm 5D} = g_A g_B / \sqrt{g_A^2+g_B^2}$.
The 4D $SU(2)_L$ and $U(1)_Y$ gauge coupling constants are given  by
$g_w ={g_A}/{\sqrt{L}}$ and $g_Y ={g_Y^{\rm 5D}}/{\sqrt{L}}$.
The bare weak mixing angle $\bar \theta_W^0$ determined by the ratio of the gauge couplings  is given by 
\begin{align}
\sin \bar \theta_W^0 = \frac{g_Y}{\sqrt{g_w^2 +g_Y^2}}   = \frac{g_B}{\sqrt{g_A^2 + 2 g_B^2}} ~.
\label{bareWangle}
\end{align}
As is seen below, the mixing angle determined from  the ratio $m_W/m_Z$ 
slightly differs from the one defined in (\ref{bareWangle}) even at the tree level in GHU in the RS space.

The 4D Higgs boson field $\Phi_H (x)$ appears as a part of $A_y^{SO(5)}$.  
$A_z^{SO(5)} = (kz)^{-1} A_y^{SO(5)}$ ($1 \le z \le z_L$)  in the tensor representation is expanded as
\begin{align}
A_z^{(j5)} (x, z) &= \frac{1}{\sqrt{k}} \, \phi_j (x) u_H (z) + \cdots ,~~
u_H (z) = \sqrt{ \frac{2}{z_L^2 -1} } \, z ~, \cr
\noalign{\kern 5pt}
\Phi_H (x) &= \frac{1}{\sqrt{2}} \begin{pmatrix} \phi_2 + i \phi_1 \cr \phi_4 - i\phi_3 \end{pmatrix} .
\label{4dHiggs}
\end{align}
$\Phi_H$ develops nonvanishing expectation value at the quantum level by the Hosotani mechanism.
Without loss of generality we take $\la \phi_1 \ra , \la \phi_2 \ra , \la \phi_3 \ra  =0$ and  $\la \phi_4 \ra \not= 0$.
The AB phase $\theta_H$ in the fifth dimension is given by 
\begin{align}
\hat W &= P \exp \bigg\{ i g_A \int_{-L}^L dy \, \la A_y^{SO(5)}  \ra \bigg\}  
=  \exp \Big\{ i  \theta_H  \cdot 2 T^{(45)} \Big\} ~ .
\label{ABphase1}
\end{align}
In terms of $\theta_H$, $A_z^{(45)} $ is expanded as 
\begin{align}
&A_z^{(45)} (x, z) = \frac{1}{\sqrt{k}} \big\{ \theta_H f_H + H(x) \big\} \, u_H(z) + \cdots , \cr
\noalign{\kern 5pt}
&f_H = \frac{2}{g_A} \sqrt{ \frac{k}{z_L^2 -1}} = \frac{2}{g_w} \sqrt{ \frac{k}{L(z_L^2 -1)}} ~.
\label{ABphase3}
\end{align}
4D neutral Higgs field $H(x)$ is the fluctuation mode of the AB phase $\theta_H$.

The AB phase $\theta_H$ plays an important role in GHU.
The value of  $\theta_H$ is determined by the location of the absolute minimum of the
effective potential $V_\eff (\theta_H)$, and the Higgs boson mass $m_H$ is given by
$m_H^2 = f_H^{-2} d^2 V_\eff (\theta_H)/d \theta_H^2 \big|_{\rm min}$.
With the KK mass scale $m_\KK$ given, the allowed range of  $\theta_H$ is constrained
to reproduce the  Higgs boson and top quark masses and also to be consistent
with the current observations at low energies.

\section{The $W$ boson mass} 

In the SM the Fermi constant $G_\mu$ determined from the $\mu$-decay is  given by\cite{Sirlin1980}
\begin{align}
\frac{G_\mu}{\sqrt{2}} &= \frac{\pi \alpha}{2 s_W^2} \, \frac{1}{m_W^2}
\, ( 1 + \Delta r_\SM^{\rm loop} ) ~,   \label{WmassSM1} \\
\noalign{\kern 5pt}
s_W^2  &= 1 - \frac{m_W^2}{m_Z^2} ~,
\label{WmassSM2}
\end{align}
where 
$\alpha^{-1} = 137.035999084 (21)$, $G_\mu = 1.1663788 (6) \times 10^{-5}\, {\rm GeV}^{-2}$ and
$m_Z = 91.1876 (21)\,$ GeV.\cite{PDG2022}
$\Delta r_\SM^{\rm loop} $ represents the sum of all loop corrections, which depends on 
$\alpha$,  $m_W$, $m_Z$,  $m_H$, strong gauge coupling constant, and masses of quarks and leptons.
Combining Eqs.\ (\ref{WmassSM1}) and (\ref{WmassSM2}), one can write the $W$ boson mass in the SM as
\begin{align}
&m_W^\SM = \frac{m_Z}{\sqrt{2}}  \left[  1 + 
\sqrt{ 1 - \frac{4 \pi \alpha (1 + \Delta r_\SM^{\rm loop} )}{\sqrt{2} \,  G_\mu m_Z^2} } ~\right]^{1/2} .
\label{WmassSM3}
\end{align}
At the tree level $\Delta r_\SM^{\rm loop} = 0$ so that $m_W^\SM |_{\rm tree} = 80.9387\,$GeV, which
is much larger than the observed $W$ mass.
Significant efforts have been made to evaluate  $\Delta r_\SM^{\rm loop}$.
At the moment the estimated value is $\Delta r_\SM^{\rm loop} \simeq 0.0383 \pm 0.0004$ and
 $m_W^\SM \simeq 80.354 \pm 0.007\,$GeV.\cite{Haller2022, Awramik2021,deBlas2022}

In GHU both of the relations (\ref{WmassSM1}) and (\ref{WmassSM2}) are modified, even at the tree level.
With given $\theta_H$ 
the masses of $W$ and $Z$, $m_W= k \lambda_W$ and $m_Z= k \lambda_Z$, satisfy \cite{GUTinspired2019a}
\begin{align}
& 2S(1;\lambda_{W}, z_L )C'(1;\lambda_{W} , z_L)+\lambda_{W} \sin^2\theta_H =  0 ~,   \label{Wspectrum1} \\
\noalign{\kern 5pt}
& 2S(1;\lambda_{Z} , z_L)C'(1;\lambda_{Z} , z_L)+ \frac{\lambda_{Z} \sin^2\theta_H}{1 - \sin^2 \theta_W^0}  =  0 ~,
\label{Zspectrum1}
\end{align}
where  the functions $C(z;\lambda, z_L)$ and $S(z; \lambda, z_L)$ are expressed in terms of Bessel functions, 
as given in Eq.\ (\ref{functionA1}).
At the tree level $\sin^2 \theta_W^0$ in (\ref{Zspectrum1}) is  equal to $\sin^2 \bar \theta_W^0$ given in (\ref{bareWangle}).
With the  orbifold  boundary condition (\ref{BC-gauge1}),  physical $m_W$ and $m_Z$ and $\sin^2 \theta_W^0$ specified, 
the wave functions of $W$ and $Z$ are determined with the conditions  (\ref{Wspectrum1}) and (\ref{Zspectrum1}).
The boundary condition does not change by radiative corrections.
In other words,  $\sin^2  \theta_W^0$ appearing in  (\ref{Zspectrum1}) is
the bare weak mixing angle in the on-shell scheme in GHU, corresponding to $s_W^2$, (\ref{WmassSM2}), in the SM.
$m_Z$ is one of the input parameters.   With $m_\KK = \pi k /(z_L -1)$ specified, the relation (\ref{Zspectrum1})
fixes the value of $z_L$ and $k$.  Then the relation (\ref{Wspectrum1}) determines $\lambda_W$ and $m_W$.
The ratio $m_W/m_Z$ thus determined is slightly different from $\cos \theta_W^0$.
For $\theta_H=0.1$, $m_\KK = 13\,$TeV, $\sin^2 \theta_W^0 = 0.2227$, for instance, one finds
$m_W - m_Z \cos\theta_W^0 = - 1.59\,$MeV.

The value of $\sin^2 \theta_W^0$ needs to be determined self-consistently such that the observed $G_\mu$
be reproduced.  In the GUT inspired GHU $\alpha$, $G_\mu$, $m_Z$, strong gauge coupling constant,  
masses of quarks and leptons, and $m_H$ are input parameters.  
With given $\sin^2 \theta_W^0$, $\theta_H$ and $m_\KK$,  one can evaluate the mass spectra of 
KK gauge bosons and their couplings to leptons.  The $\mu$-decay proceeds, at the tree level, 
by emitting not only $W=W^{(0)}$, but also $W^{(n)}$ and $W_R^{(n)}$ ($n \ge 1$).  Here $W_R^{(n)}$
are gauge bosons in $SU(2)_R$ of $SO(4)=SU(2)_L \times SU(2)_R \subset SO(5)$.
Hence the relation (\ref{WmassSM1}) is replaced by
\begin{align}
\frac{G_\mu}{\sqrt{2}} &= \frac{\pi \alpha}{2 \sin^2 \theta_W^0} \,
 \frac{\hat g^{W^{(0)}}_{\mu\nu_\mu, L}  \hat g^{W^{(0)}}_{e \nu_e, L}   }{m_{W^{(0)}}^2}
\, ( 1 + \Delta r_G) \, ( 1 + \Delta r_\GHU^{\rm loop} ) ,   \cr
\noalign{\kern 5pt}
 \Delta r_G &= \frac{1}{\hat g^{W^{(0)}}_{\mu\nu_\mu, L}  \hat g^{W^{(0)}}_{e \nu_e, L} }   \sum_{n=1}^\infty 
 \bigg\{ \hat g^{W^{(n)}}_{\mu\nu_\mu, L}  \hat g^{W^{(n)}}_{e \nu_e, L}  \Big[ \frac{m_{W^{(0)}}}{m_{W^{(n)}}} \Big]^2
 +  \hat g^{W_R^{(n)}}_{\mu\nu_\mu, L}  \hat g^{W_R^{(n)}}_{e \nu_e, L}  \Big[ \frac{m_{W^{(0)}}}{m_{W_R^{(n)}}} \Big]^2 \bigg\} ,
\label{GHUfermi1}
\end{align}
where the coupling of $W^{(n)}$ to $e \nu_e$, for instance, is given by
$(g_w/\sqrt{2})  W^{(n)}_\mu \big\{ \hat g^{W^{(n)}}_{e \nu_e, L} \,  \bar \nu_{e,L} \gamma^\mu e_L +
 \hat g^{W^{(n)}}_{e \nu_e, R} \, \bar \nu_{e,R} \gamma^\mu e_R  \big\}$.
The right-handed couplings are very small ($| \hat g^{W^{(n)}}_{e \nu_e, R} | < 10^{-19}$ etc.), 
and have been omitted in the expression for $ \Delta r_G$ in (\ref{GHUfermi1}).
$ \Delta r_\GHU^{\rm loop} $ represents the sum of loop corrections.
In GHU the $W$ boson mass $m_W=m_{W^{(0)}}$ is determined by solving (\ref{Wspectrum1}) and (\ref{GHUfermi1})
simultaneously.

The mass spectra $\{  m_{W^{(n)}} = k \lambda_{W^{(n)}} \}$ and $\{  m_{W_R^{(n)}} = k \lambda_{W_R^{(n)}} \}$
are determined by
\begin{align}
& 2S(1;\lambda_{W^{(n)}}, z_L )C'(1;\lambda_{W^{(n)}} , z_L)+\lambda_{W^{(n)}} \sin^2\theta_H =  0 ~,   \label{Wspectrum2} \\
&C(1;\lambda_{W_R^{(n)}}  , z_L) =  0 ~,
\label{WRspectrum}
\end{align}
respectively.  Wave functions of gauge and fermion fields are also determined, with which gauge couplings among them 
are evaluated.  (See \cite{sumrule2023} and \cite{GHUfiniteT2021} for details.)
The values of $\big\{ m_{W^{(n)}},  \hat g^{W^{(n)}}_{e \nu_e, L} ,  \hat g^{W^{(n)}}_{\mu \nu_\mu, L}  \big\}$ and 
$\big\{ m_{W_R^{(n)}},  \hat g^{W_R^{(n)}}_{e \nu_e, L} , \hat g^{W_R^{(n)}}_{\mu \nu_\mu, L} \big\}$ are tabulated 
in Table \ref{Table:mWn} and Table  \ref{Table:mWRn} for $m_\KK = 13\,$TeV and $\theta_H = 0.10$, respectively. 
One sees that the $W^{(1)}$ mode has large couplings $\hat g^{W^{(1)}}_{e \nu_e, L} \sim 5.721$ and 
$ \hat g^{W^{(1)}}_{\mu \nu_\mu, L} \sim 5.446$, giving an appreciable correction to $G_\mu$.
The infinite sum in $\Delta r_G$ in (\ref{GHUfermi1}) is seen to rapidly converge.

$\Delta r_\GHU^{\rm loop}$ in (\ref{GHUfermi1}) represents radiative corrections.
KK excited modes of  gauge bosons, leptons, and quarks give little contributions to $\Delta r_\GHU^{\rm loop}$,
as their masses are of $O(m_\KK)$ and $m_\mu/m_\KK \ll 1$.
Only SM particles give relevant contributions to $\Delta r_\GHU^{\rm loop}$, and the couplings among
the SM particles are nearly the same as in the SM.  (For instance, $\hat g^{W^{(0)}}_{e \nu_e, L}  = 0.997649$
as shown in Table \ref{Table:mWn}.). 
In ref.\ \cite{sumrule2023} oblique corrections to the $W$, $Z$ and $\gamma$ propagators, namely 
Peskin-Takeuchi $S$, $T$, $U$ parameters,\cite{PeskinTakeuchi}
due to the KK modes of quarks and leptons  have been evaluated.
It has been shown that as a result of the coupling sum rules the oblique corrections are small.  
The contributions of KK excited modes of quark-lepton multiplets to $S$, $T$, and $U$ per KK level are
$\delta S \sim 0.002$, $\delta T \sim 0.02$ and $\delta U \sim 10^{-5}$.\cite{sumrule2023}
Contributions to oblique parameters of $W^{(n)}$, $W_R^{(n)}$ ($n \ge 1$) also need to be
taken into account for complete analysis.  We expect that they are  small as well.
It is reasonable to approximate $\Delta r_\GHU^{\rm loop}$  by $\Delta r_\SM^{\rm loop}$.
In the evaluation below we take $\Delta r_\GHU^{\rm loop} = \Delta r_\SM^{\rm loop} = 0.0383$.

\begin{table}[tbh]
{
\renewcommand{\arraystretch}{1.2}
\begin{center}
\caption{The masses $m_{W^{(n)}}$ and couplings $\hat g^{W^{(n)}}_{e \nu_e, L} ,  \hat g^{W^{(n)}}_{\mu \nu_\mu, L}$
$(n = 0, 1, \cdots,9)$  are shown for $m_\KK = 13\,$TeV and $\theta_H = 0.10$ with $\sin^2 \theta_W^0 = 0.22266$.  
Right-handed couplings are very small;  $| \hat g^{W^{(n)}}_{e \nu_e, R} | <  2 \times 10^{-20}$ and 
$| \hat g^{W^{(n)}}_{\mu \nu_\mu, R} | <  8 \times 10^{-19}$  for $n \le 14$.
 }
\vskip 10pt
\begin{tabular}{|c|c|c|c|c|c|}
\hline
$n$ & 
$\genfrac{}{}{0pt}{}{\displaystyle \strut m_{W^{(n)}}}{\hbox{\small \rm [GeV]}}$
& $\myfrac{m_{W^{(n)}}}{m_\KK}$  & $\myfrac{m_{W^{(0)}}}{m_{W^{(n)}}}$ 
&  $\hat g^{W^{(n)}}_{e \nu_e, L} $ & $ \hat g^{W^{(n)}}_{\mu \nu_\mu, L}$\\
\hline
$0$ &80.396 &0.0062 &1  &0.997649 &0.997646 \\
\hline
$1$ &10199. &0.7845 &$7.88 \times 10^{-3}$ &5.72126 &5.44645 \\
\hline
$2$ &15857. &1.2198 &$5.07 \times 10^{-3}$ &0.01858 &0.01641\\
\hline
$3$ &23102. &1.7771 &$3.48 \times 10^{-3}$ &2.26066 &1.72755 \\
\hline
$4$ &29032. &2.2332 &$2.77 \times 10^{-3}$ &0.00607 &0.00421\\
\hline
$5$ &36074. &2.7749 &$2.23 \times 10^{-3}$ &0.82175 &0.59526 \\
\hline
$6$ &42099. &3.2384 &$1.91 \times 10^{-3}$ &0.00291 &0.00226 \\
\hline
$7$ &49062. &3.7740 &$1.64 \times 10^{-3}$ &0.48331 &0.36385 \\
\hline
$8$ &55135. &4.2412 &$1.46 \times 10^{-3}$ &0.00173 &0.00123 \\
\hline
$9$ &62056. &4.7735 &$1.30 \times 10^{-3}$ &0.28815 &0.20853\\
\hline
\end{tabular}
\hskip 1pt
\label{Table:mWn}
\end{center}
}
\end{table}

\begin{table}[tbh]
{
\renewcommand{\arraystretch}{1.2}
\begin{center}
\caption{The masses $m_{W_R^{(n)}}$ and couplings  $\hat g^{W_R^{(n)}}_{e \nu_e, L} ,  \hat g^{W_R^{(n)}}_{\mu \nu_\mu, L}$ 
$(n = 1, 2, \cdots,5)$ are shown for $m_\KK = 13\,$TeV and $\theta_H = 0.10$ with $\sin^2 \theta_W^0 = 0.22266$.  
Right-handed couplings are very small;  $| \hat g^{W_R^{(n)}}_{e \nu_e, R} | <  2 \times 10^{-19}$ 
and $| \hat g^{W_R^{(n)}}_{\mu \nu_\mu, R} | <  2 \times 10^{-17}$ for $n \le  8$.
 }
\vskip 10pt
\begin{tabular}{|c|c|c|c|c|c|}
\hline
$n$ & 
$\genfrac{}{}{0pt}{}{\displaystyle \strut m_{W_R^{(n)}}}{\hbox{\small \rm [GeV]}}$
& $\myfrac{m_{W_R^{(n)}}}{m_\KK}$& $\myfrac{m_{W^{(0)}}}{m_{W_R^{(n)}}}$ 
&  $\hat g^{W_R^{(n)}}_{e \nu_e, L} $ & $ \hat g^{W_R^{(n)}}_{\mu \nu_\mu, L}$\\
\hline
$1$ &9951. &0.7655 &$8.08 \times 10^{-3}$ &0.01449 &0.01382 \\
\hline
$2$ &22842. &1.7571 &$3.52 \times 10^{-3}$ &0.00579 &0.00445 \\
\hline
$3$ &35809. &2.7546 &$2.25 \times 10^{-3}$ &0.00209 &0.00151\\
\hline
$4$ &48794. &3.7534 &$1.65 \times 10^{-3}$ &0.00122 &0.00092 \\
\hline
$5$ &61785. &4.7527 &$1.30 \times 10^{-3}$ &0.00073 &0.00053\\
\hline
\end{tabular}
\label{Table:mWRn}
\end{center}
}
\end{table}

Now one can evaluate $m_W$ for given $m_\KK$ and $\theta_H$.
Specify a tentative value for $\sin^2 \theta_W^0$, with which one  determines 
$m_W$ from (\ref{Wspectrum1}), and also from (\ref{GHUfermi1}).  
The two values generally differ from each other.  
We adjust $\sin^2 \theta_W^0$ such that these two values for $m_W$ coincide.
In this manner $m_W$ and $\sin^2 \theta_W^0$ are determined to satisfy  (\ref{Wspectrum1}) and (\ref{GHUfermi1})
simultaneously.
For $m_\KK = 13\,$TeV and $\theta_H = 0.10$, for instance, we find that
$m_W = 80.396\,$GeV, $\sin^2 \theta_W^0 = 0.22266$,
 $\hat g^{W^{(0)}}_{e \nu_e, L} \hat g^{W^{(0)}}_{\mu \nu_\mu, L} = 0.99530$ and $\Delta r_G = 0.0020$.

\begin{figure}[tbh]
\centering
\includegraphics[height=60mm]{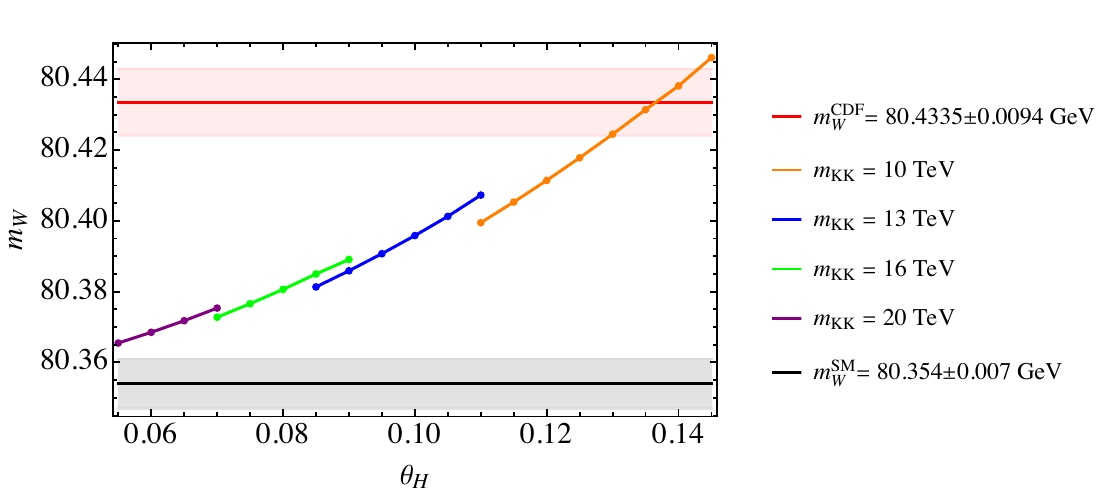}
\caption{The $W$ boson mass $m_W$ in GHU is plotted as a function of $\theta_H$ with various $m_\KK$.
The constraint $m_\KK \gtrsim 13\,$TeV is obtained from  the experimental data at LHC.\cite{Funatsu2022a}
The predicted $m_W$ in GHU for $13\, {\rm TeV} \le m_\KK \le 20\, {\rm TeV}$ lies between
$m_W^\SM$ and $m_W^{\rm CDF}$.}
\label{fig:mW}
\end{figure}

In Fig.~\ref{fig:mW} the predicted values for $m_W$ are plotted for various values of  $m_\KK$ and $\theta_H$.
It is seen that  $m_W$ in GHU becomes larger than $m_W^\SM$ in the SM, but is smaller than $m_W^{\rm CDF}$
for $13\,{\rm TeV} \le m_\KK \le 20\,{\rm TeV}$.
The GUT inspired GHU in the RS space naturally predicts the $W$ boson mass well above $m_W^\SM$.
We note that  an uncertainty  $\Delta m_W$ of about $7\,$MeV is expected as in the SM for an uncertainty  in $m_W^\SM$.

\begin{figure}[htb]
\centering
\includegraphics[height=60mm]{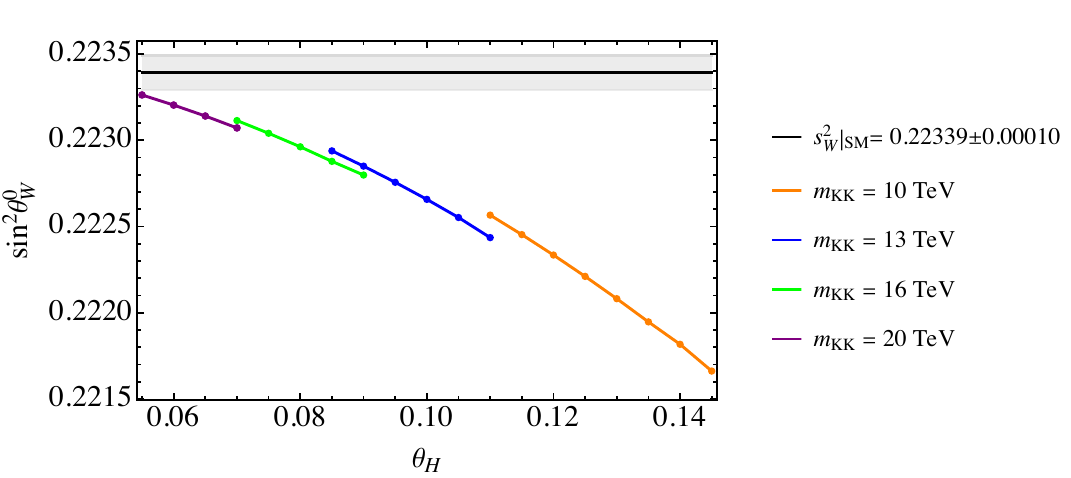}
\caption{$\sin^2 \theta_W^0$ in the on-shell scheme is plotted as a function of $\theta_H$ with various $m_\KK$.
$s_W^2 |_{\SM}  = 1 - (m_W^2/m_Z^2) = 0.22339 \pm 0.00010$ is the value  in the on-shell scheme
in the SM, listed in Table 10.2 of Ref.\ \cite{PDG2022}.
}   
\label{fig:sin2thetaW0}
\end{figure}

\begin{figure}[htb]
\centering
\includegraphics[height=60mm]{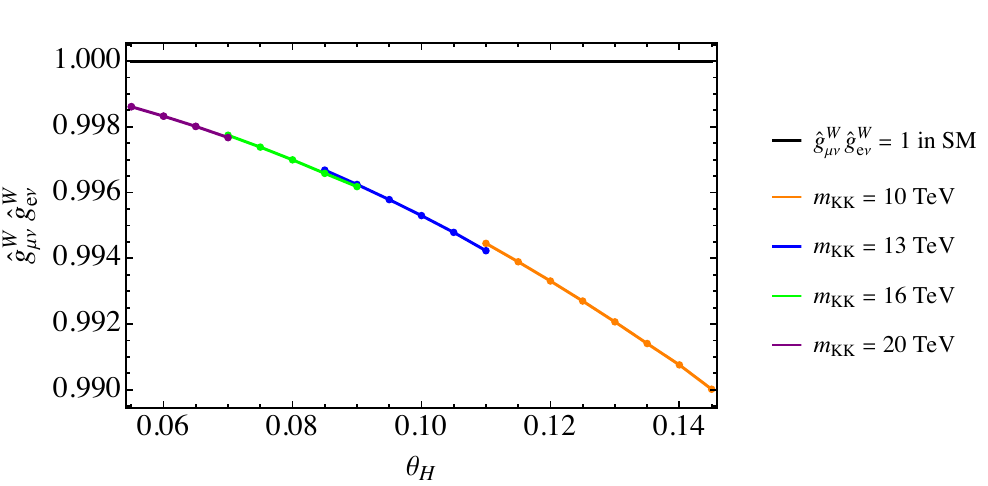}
\caption{The product of the $W$ couplings of $\mu$ and $e$ normalized by the SM coupling $g_w/\sqrt{2}$, 
$\hat g^{W}_{\mu\nu} \,  \hat g^{W}_{e \nu} \equiv \hat g^{W^{(0)}}_{\mu\nu_\mu, L} \,  \hat g^{W^{(0)}}_{e \nu_e, L} $,
is plotted as a function of $\theta_H$ with various $m_\KK$.}   
\label{fig:gg00}
\end{figure}

\begin{figure}[hbt]
\centering
\includegraphics[height=60mm]{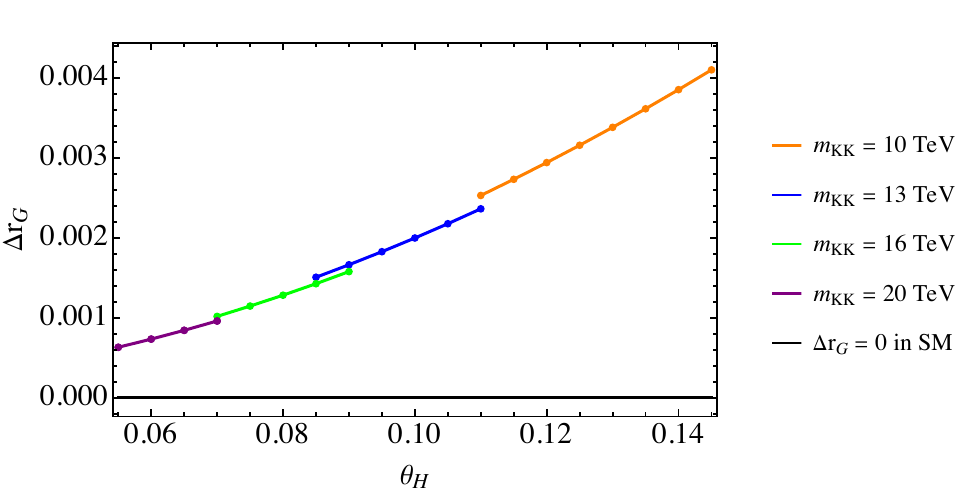}
\caption{$\Delta r_G$  in Eq.\ (\ref{GHUfermi1})  is plotted as a function of $\theta_H$ with various $m_\KK$.
$\Delta r_G = 0$ in the SM.}
\label{fig:DelrG}
\end{figure}

In Figs.~\ref{fig:sin2thetaW0}, \ref{fig:gg00} and \ref{fig:DelrG},  $\sin^2 \theta_W^0$,  
$\hat g^{W^{(0)}}_{e \nu_e, L} \hat g^{W^{(0)}}_{\mu \nu_\mu, L} $ and $\Delta r_G$ are displayed, respectively.
It is observed that these quantities are mostly determined by the value of $\theta_H$. 
It is seen in Fig.~\ref{fig:sin2thetaW0} that $\sin^2 \theta_W^0$ in the on-shell scheme in GHU approaches to
$s_W^2 |_{\SM}  = 1 - (m_W^2/m_Z^2)$ in the on-shell scheme in the SM as  $m_\KK$ becomes larger.
Similarly $\hat g^{W^{(0)}}_{e \nu_e, L} \hat g^{W^{(0)}}_{\mu \nu_\mu, L} $ and $\Delta r_G$ 
also approach  to those in the SM as  $m_\KK$ becomes large.

To compare the result in GHU to that in the SM, let us rewrite the formula (\ref{GHUfermi1}) in the form
\begin{align}
\frac{G_\mu}{\sqrt{2}} &=C \,  \frac{\pi \alpha}{2s_{W, \GHU}^2} \,
 \frac{1 }{m_{W^{(0)}}^2} \, ( 1 + \Delta r_\GHU^{\rm loop} ) ,   \cr
\noalign{\kern 5pt}
C &= \frac{s_{W,\GHU}^2}{ \sin^2 \theta_W^0} \, \hat g^{W^{(0)}}_{\mu\nu_\mu, L}  \hat g^{W^{(0)}}_{e \nu_e, L} 
 \,  ( 1 + \Delta r_G) ~, 
\label{GHUfermi2}
\end{align}
where $s_{W,\GHU}^2 = 1 - ( m_{W^{(0)}}^2/m_Z^2)$.
There are three factors in $C$.  For $m_\KK = 13\,$TeV and $\theta_H = 0.10$ one finds
$s_{W,\GHU}^2 / \sin^2 \theta_W^0 = 1.00014$, $\hat g^{W^{(0)}}_{\mu\nu_\mu, L}  \hat g^{W^{(0)}}_{e \nu_e, L} =0.99530$,
$1 + \Delta r_G = 1.0020$ and therefore $C = 0.997425 < 1$.
In other words the effective $\Delta r_\GHU ^\eff$ defined by $1 + \Delta r_\GHU^\eff = C (1 + \Delta r_\GHU^{\rm loop})$
becomes smaller than $ \Delta r_\GHU^{\rm loop} \sim  \Delta r_\SM^{\rm loop}$, which in turn makes $m_W^{(0)}$
larger than $m_W^\SM$.

\begin{figure}[bht]
\centering
\includegraphics[height=60mm]{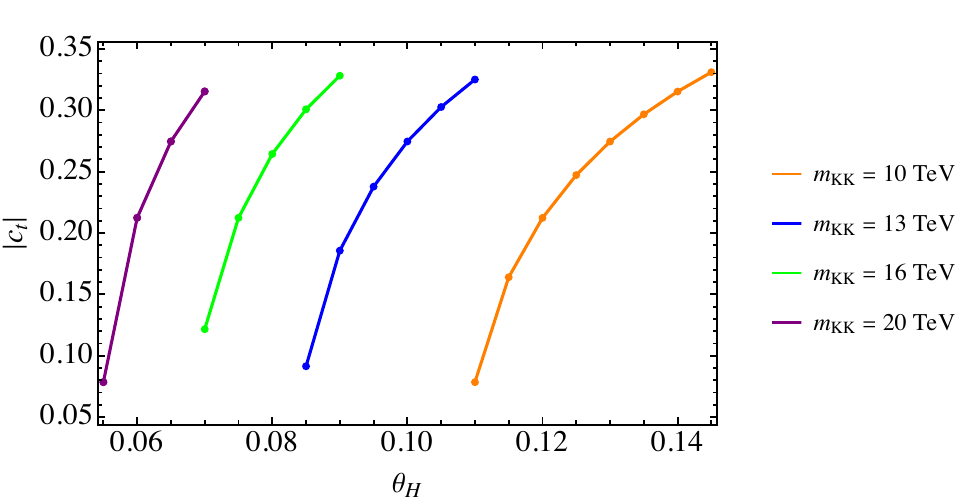}
\caption{The absolute value of the bulk mass parameter  of the top quark multiplet field, $c_t$, is plotted as a function 
of $\theta_H$ with various $m_\KK$.}
\label{fig:ctop}
\end{figure}

In Figs.~\ref{fig:mW}-\ref{fig:DelrG} the range of $\theta_H$ is
restricted to $\theta_H^{\rm min} \le  \theta_H \le  \theta_H^{\rm max}$,  once $m_\KK$ is specified.
For  $\theta_H < \theta_H^{\rm min}$ the top quark mass $m_t$ cannot be reproduced. 
The top quark mass $m_t = k \lambda_t$ is determined by
\begin{align}
&S_L(1; \lambda_t, c_t; z_L) S_R(1; \lambda_t, c_t; z_L)  + \sin^2 \frac{\theta_H}{2} = 0 
\label{topmass1}
\end{align}
where $c_t$ is the bulk mass parameter of the top quark field and $S_{L/R} (z; \lambda. c, z_L)$
is given by Eq.~(\ref{functionA2}). 
Eq.~(\ref{topmass1}) is invariant under  $c_t \go - c_t$.
With given $\theta_H$, $z_L$, and $\lambda_t$, a solution for $|c_t|$ exists only for $\theta_H \ge \theta_H^{\rm min}$.
($c_t =0$ for $\theta_H = \theta_H^{\rm min}$).
In Fig.~\ref{fig:ctop},  $|c_t|$ as a function of $\theta_H$ with various  $m_\KK$ is displayed.
We note that the bulk mass parameters of the other up-type quarks and charged leptons are determined
by the same form of the equations as Eq.~(\ref{topmass1}).   
For $m_\KK=13\,$TeV and $\theta_H=0.1$, for instance, one finds
$(c_u, c_c, c_t) = ( -0.863,  -0.722, -0.275)$ and $( c_e,  c_\mu, c_\tau) = ( -1.012, -0.796, -0.677)$.
The mass hierarchy of quarks and leptons is naturally explained by $O(1)$ bulk mass parameters in GHU.
Only the top quark field has $|c_t| < \onehalf$ as $m_t > m_W$.
In the GUT inspired GHU negative values for the bulk mass parameters for quark and lepton multiplet fields have 
been adopted.  Positive bulk mass parameters would lead to additional MeV scale neutrinos in the lepton sector and
additional very  light KK modes of down-type quarks in the first and second generations,
the latter of which conflicts with the observation.\cite{GUTinspired2019a}

\begin{figure}[htb]
\centering
\includegraphics[height=60mm]{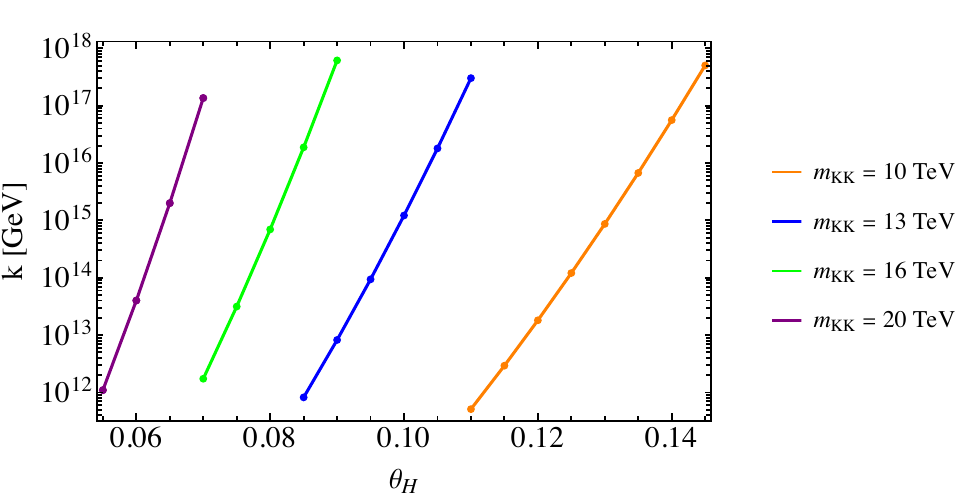}
\caption{The parameter $k$ in the AdS curvature of the RS warped space $\Lambda = - 6 k^2$ is depicted  
as a function of $\theta_H$ with various $m_\KK$.  As $k$ approaches  the Planck mass scale, 
the $\mu$-$e$ universality starts to break down.}
\label{fig:AdSk}
\end{figure}

The value of  $\theta_H$ is also constrained by $\theta_H < \theta_H^{\rm max}$.
It turns out that for $\theta_H > \theta_H^{\rm max}$ the $\mu$-$e$ universality is spoiled 
particularly in the $Z$ couplings of the right-handed $e$ and $\mu$ at the $m_Z$ scale.
For instance, with $m_\KK = 13\,$TeV, the $\mu$-$e$ universality holds within an error of $3 \times 10^{-6}$
for $0.09 \le \theta_H \le 0.11$, but the universality breaks with a magnitude $3 \times 10^{-3}$ for $\theta_H = 0.115$.
It seems to  be related to the fact that the AdS curvature ($\Lambda = - 6 k^2$) becomes large.
$8 \times 10^{12}\,{\rm GeV} \le k \le  3 \times 10^{17}\,{\rm GeV}$  for $0.09 \le \theta_H \le 0.11$, 
whereas $k = 6 \times 10^{18}\,$GeV for $\theta_H = 0.115$, the value of $k$ getting close 
to the Planck mass scale. See Fig.~\ref{fig:AdSk}. 
We also note that another constraint $m_\KK \gtrsim 13\,$TeV is obtained from the experimental data 
at LHC.\cite{Funatsu2022a}

\begin{figure}[htb]
\centering
\includegraphics[height=60mm]{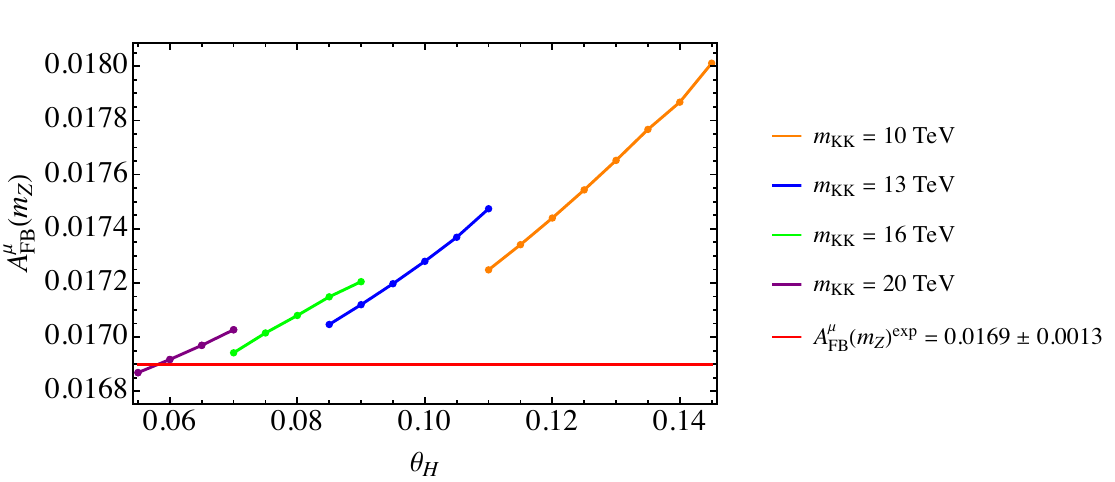}
\caption{The forward-backward asymmetry in the process $e^- e^+ \go \mu^- \mu^+$ at the $Z$ pole 
$A_{\rm FB}^\mu (m_Z)$ is plotted as a function of $\theta_H$ with various $m_\KK$.
$A_{\rm FB}^\mu (m_Z)$ for $13\, {\rm TeV} \le m_\KK \le 20\,{\rm TeV}$ is consistent with
the observed value $A_{\rm FB}^\mu (m_Z)^{\rm exp} = 0.0169 \pm 0.0013$.}
\label{fig:AFB}
\end{figure}

Once $m_\KK$ and $\theta_H$ are specified, $\sin^2 \theta_W^0$ in the on-shell scheme in GHU  
is determined as described above. 
Now we estimate the forward-backward asymmetry $A_{\rm FB}^\mu$ in the $e^- e^+ \go \mu^- \mu^+$
process at the $m_Z$ pole.  For this end we need to know  $\sin^2 \bar \theta_W^0 (m_Z)$ at the $m_Z$ scale
corresponding to $\hat s_Z^2 = \sin^2 \hat \theta_W (m_Z)$ in the $\overline{\rm MS}$ scheme in the SM.\cite{PDG2022}
In relating $\sin^2 \theta_W^0$ to $\sin^2 \bar \theta_W^0 (m_Z)$, 
only SM particles are relevant as all KK modes are very heavy ($m_\KK \gg m_Z$).
Further the couplings among SM particles in GHU are nearly the same as those in the SM.
Therefore it is reasonable to approximate as 
$\sin^2 \bar \theta_W^0 (m_Z) \sim K_\SM \sin^2 \theta_W^0$
where $K_\SM  = \hat s_Z^2 /s_W^2 \sim 0.23122/0.22339$.\cite{PDG2022}
With this $\sin^2 \bar \theta_W^0 (m_Z)$ one evaluates the $Z$ couplings of $e$ and $\mu$ at the $m_Z$ scale, 
from which $A_{\rm FB}^\mu (m_Z)$ is determined.  
The result is plotted in Fig.~\ref{fig:AFB}.  It is seen that the predicted values in GHU are consistent
with the experimental data  $A_{\rm FB}^{\mu} (m_Z)^{\rm exp} = 0.0169 \pm 0.0013$ for
 $13\, {\rm TeV} \le m_\KK \le 20\, {\rm TeV}$.

\section{Summary} 

In this paper we have evaluated the $W$ boson mass in the GUT inspired GHU in the RS space.
With the KK mass scale $m_\KK$ specified, the allowed range of the AB phase $\theta_H$ is
constrained as $\theta_H^{\rm min} \le \theta_H \le \theta_H^{\rm max}$.  
For $m_\KK = 13\,$TeV, for instance, $0.085  \lesssim  \theta_H \lesssim 0.11$, 
and the predicted $m_W$ is $80.381\,{\rm GeV}  \lesssim m_W \lesssim 80.407\,{\rm GeV}$.
The result for other values of $m_\KK$ is depicted in Fig.~\ref{fig:mW}.
For $13 \, {\rm TeV} \le m_\KK \le 20\, {\rm TeV}$, the predicted $m_W$ lies between
the SM value  $m_W^{\rm SM} = 80.354 \pm 0.007\,$ and the CDF value
$m_W^{\rm CDF} = 80.4335 \pm 0.0094\,$GeV.

In the GUT inspired GHU the value of $m_W$ is determined by (\ref{Wspectrum1}) and  (\ref{GHUfermi1}).
In addition to $W= W^{(0)}$ the KK excited modes $W^{(n)}$ and $W_R^{(n)}$ ($n \ge 1$) mediate
the $\mu$ decay at the tree level.  The $W$ couplings of $e$ and $\mu$, $\hat g^{W^{(0)}}_{e \nu_e, L}$ and
$\hat g^{W^{(0)}}_{\mu\nu_\mu, L}$, become slightly smaller than those in the SM,
which leads to a larger value for $m_W$ than that in the SM.
It is curious that $m_W$ is mostly determined by the value of the AB phase $\theta_H$ as
seen in Fig.~\ref{fig:mW}.

It is extremely important to definitively determine $m_W$ by experiments.
The $W^+ W^-$ production process in the $e^+ e^-$ collisions near the threshold should
give indispensable information on $m_W$.  
Once $m_W$ is determined, the values of $m_\KK$ and $\theta_H$ in GHU can be severely constrained.

\section*{Acknowledgment}

This work was supported in part
by the Ministry of Science and Technology of Taiwan under
Grant No. MOST-111-2811-M-002-047-MY2 (N.Y.).


\appendix

\section{Basis functions} 

We  summarize the basis functions used for wave functions of gauge and fermion fields.
For gauge fields we introduce
\begin{align}
 F_{\alpha, \beta}(u, v) &\equiv J_\alpha(u) Y_\beta(v) - Y_\alpha(u) J_\beta(v) ~, \cr
\noalign{\kern 5pt}
 C(z; \lambda, z_L) &= \frac{\pi}{2} \lambda z z_L F_{1,0}(\lambda z, \lambda z_L) ~,  \cr
 S(z; \lambda, z_L) &= -\frac{\pi}{2} \lambda  z F_{1,1}(\lambda z, \lambda z_L) ~, \cr
 C^\prime (z; \lambda, z_L) &= \frac{\pi}{2} \lambda^2 z z_L F_{0,0}(\lambda z, \lambda z_L) ~,  \cr
S^\prime (z; \lambda, z_L) &= -\frac{\pi}{2} \lambda^2 z  F_{0,1}(\lambda z, \lambda z_L)~, 
\label{functionA1}
\end{align}
where $J_\alpha (u)$ and $Y_\alpha (u)$ are Bessel functions of  the first and second kind.
They satisfy
\begin{align}
&- z \frac{d}{dz} \frac{1}{z} \frac{d}{dz} \begin{pmatrix} C \cr S \end{pmatrix} 
= \lambda^{2} \begin{pmatrix} C \cr S \end{pmatrix} ~,  
\label{relationA1}
\end{align}
with the boundary conditions $C(z_{L} ; \lambda, z_L)  = z_{L}$, $C' (z_{L} ; \lambda, z_L)  =S(z_{L} ; \lambda, z_L)  = 0 $, 
$S' (z_{L} ; \lambda, z_L)  = \lambda$, and $CS' - S C' = \lambda z$.

For fermion fields with a bulk mass parameter $c$, we define 
\begin{align}
\begin{pmatrix} C_L \cr S_L \end{pmatrix} (z; \lambda,c, z_L)
&= \pm \frac{\pi}{2} \lambda \sqrt{z z_L} F_{c+\frac12, c\mp\frac12}(\lambda z, \lambda z_L) ~, \cr
\begin{pmatrix} C_R \cr S_R \end{pmatrix} (z; \lambda,c, z_L)
&= \mp \frac{\pi}{2} \lambda \sqrt{z z_L} F_{c- \frac12, c\pm\frac12}(\lambda z, \lambda z_L) ~.  
\label{functionA2}
\end{align}
These functions satisfy 
\begin{align}
&D_{+} (c) \begin{pmatrix} C_{L} \cr S_{L} \end{pmatrix} = \lambda  \begin{pmatrix} S_{R} \cr C_{R} \end{pmatrix}, \cr
\noalign{\kern 5pt}
&D_{-} (c) \begin{pmatrix} C_{R} \cr S_{R} \end{pmatrix} = \lambda  \begin{pmatrix} S_{L} \cr C_{L} \end{pmatrix}, ~~
D_{\pm} (c) = \pm \frac{d}{dz} + \frac{c}{z} ~, 
\label{relationA2}
\end{align}
with the boundary conditions $C_{R/L} =1$, $S_{R/L} = 0$ at $z=z_{L} $, and 
$C_L C_R - S_L S_R=1$.

\def\jnl#1#2#3#4{{#1}{\bf #2},  #3 (#4)}

\def\Zphys{{\em Z.\ Phys.} }
\def\jssc{{\em J.\ Solid State Chem.\ }}
\def\jpsJ{{\em J.\ Phys.\ Soc.\ Japan }}
\def\ptps{{\em Prog.\ Theoret.\ Phys.\ Suppl.\ }}
\def\PTP{{\em Prog.\ Theoret.\ Phys.\  }}
\def\PTEP{{\em Prog.\ Theoret.\ Exp.\  Phys.\  }}
\def\JMP{{\em J. Math.\ Phys.} }
\def\NPB{{\em Nucl.\ Phys.} B}
\def\NP{{\em Nucl.\ Phys.} }
\def\PLB{{\it Phys.\ Lett.} B}
\def\PL{{\em Phys.\ Lett.} }
\def\PRL{\em Phys.\ Rev.\ Lett. }
\def\PRB{{\em Phys.\ Rev.} B}
\def\PRD{{\em Phys.\ Rev.} D}
\def\PRe{{\em Phys.\ Rep.} }
\def\AP{{\em Ann.\ Phys.\ (N.Y.)} }
\def\RMP{{\em Rev.\ Mod.\ Phys.} }
\def\ZPC{{\em Z.\ Phys.} C}
\def\SCI{{\em Science} }
\def\CMP{\em Comm.\ Math.\ Phys. }
\def\MPLA{{\em Mod.\ Phys.\ Lett.} A}
\def\IJMPA{{\em Int.\ J.\ Mod.\ Phys.} A}
\def\IJMPB{{\em Int.\ J.\ Mod.\ Phys.} B}
\def\EPJC{{\em Eur.\ Phys.\ J.} C}
\def\EPJP{{\em Eur.\ Phys.\ J.} Plus}
\def\PR{{\em Phys.\ Rev.} }
\def\JHEP{{\em JHEP} }
\def\JCAP{{\em JCAP} }
\def\cmp{{\em Com.\ Math.\ Phys.}}
\def\JPA{{\em J.\  Phys.} A}
\def\JPG{{\em J.\  Phys.} G}
\def\NJP{{\em New.\ J.\  Phys.} }
\def\CQG{\em Class.\ Quant.\ Grav. }
\def\ATMP{{\em Adv.\ Theoret.\ Math.\ Phys.} }
\def\ibid{{\em ibid.} }
\def\ChP{{\em Chin.Phys.}C}
\def\NCA{{\it Nuovo Cim.} A}


\renewenvironment{thebibliography}[1]
         {\begin{list}{[$\,$\arabic{enumi}$\,$]}  
         {\usecounter{enumi}\setlength{\parsep}{0pt}
          \setlength{\itemsep}{0pt}  \renewcommand{\baselinestretch}{1.2}
          \settowidth
         {\labelwidth}{#1 ~ ~}\sloppy}}{\end{list}}

\vskip 1.cm

\leftline{\Large \bf References}


\end{document}